\documentclass[pre,twocolumn,showpacs,preprintnumbers,
        amsmath,amssymb,superscriptaddress]{revtex4}
\usepackage[latin1]{inputenc}
\usepackage{graphics}
\usepackage{color}
\usepackage{dcolumn}% Align table columns on decimal point
\usepackage{graphicx}% Include figure files
\usepackage{dcolumn}% Align table columns on decimal point
\usepackage{bm}% bold math

\begin{document}

%\preprint{APS/123-QED}

\title{Dynamics in dense hard-sphere colloidal suspensions}

\author{Davide Orsi}
\affiliation{European Synchrotron Radiation Facility, B.P. 220, 38043 Grenoble, France}
\affiliation{Physics Department, University of Parma, Viale Usberti 7/A, 
Parma 43100, Italy}
\author{Andrei Fluerasu}
\email{fluerasu@bnl.gov}
\affiliation{European Synchrotron Radiation Facility, B.P. 220, 38043 Grenoble, France}
\affiliation{Brookhaven National Laboratory, NSLS-II, Upton NY 11973, USA}
\author{Abdellatif Moussa\"id}
\affiliation{European Synchrotron Radiation Facility, B.P. 220, 38043 Grenoble, France}
\affiliation{Laboratoire de Spectrom\'etrie Physique, Universit\'e Joseph Fourier, 38401 Grenoble, France}
\author{Federico Zontone}
\affiliation{European Synchrotron Radiation Facility, B.P. 220, 38043 Grenoble, France}
\author{Luigi Cristofolini}
\affiliation{Physics Department, University of Parma, Viale Usberti 7/A, 
Parma 43100, Italy}
\author{Anders Madsen}
\affiliation{European Synchrotron Radiation Facility, B.P. 220, 38043 Grenoble, France}
\affiliation{European X-Ray Free-Electron Laser, 22761 Hamburg, Germany}

\date{\today}

\begin{abstract}
The dynamic behavior of a hard-sphere colloidal suspension was studied
by X-ray Photon Correlation Spectroscopy and Small Angle X-ray Scattering
over a wide range of particle volume fractions. The short-time mobility of
the particles was found to be smaller than that of free particles even at
relatively low concentrations, showing the importance of indirect
hydrodynamic interactions. Hydrodynamic functions were derived from the
data and for moderate particle volume fractions ($\Phi\leq$~0.40) there is
a good agreement with earlier many-body theory calculations by Beenakker and
Mazur [C.W.J. Beenakker and P. Mazur, {\it Physica A} {\bf 120}, 349 (1984)].
Important discrepancies appear at higher concentrations, above
$\Phi \approx$ 0.40, where the hydrodynamic effects are overestimated
by the Beenakker-Mazur theory, but predicted accurately by an accelerated
Stokesian dynamics algorithm developed by Banchio and Brady
[A.J. Banchio and J. F. Brady, {\it J. Chem. Phys.} {\bf 118}, 10323 (2003)].
For the relaxation rates, good agreement was also found between the 
experimental data and a scaling form predicted by Mode Coupling Theory. In the 
high concentration range, with the fluid suspensions approaching the
glass transition, the long-time diffusion coefficient was compared with the 
short-time collective diffusion coefficient to verify a scaling relation 
previously proposed by Segr\`e and Pusey [P.N. Segr\`e and P.N. Pusey,
{\it Phys. Rev. Lett.} {\bf 77}, 771 (1996)]. We discuss our results in view 
of previous experimental attempts to validate this scaling law
[L. Lurio {\it et al.}, {\it Phys. Rev. Lett.} {\bf 84}, 785 (2000)].
\end{abstract}
\maketitle

\section{Introduction}

The dynamical behavior of colloidal suspensions is a very rich
research area with borderlines to many fields of fundamental research and 
important industrial applications. High density colloidal suspensions
provide, for instance, invaluable model systems for the study and
understanding of dynamics in atomic glasses. Prototypical model systems 
consist in suspensions of spherical particles with low size
polydispersity. They are stabilized against aggregation due to van der Waals
attractive forces by coating the surface with a short-chained polymer
(steric stabilization) or with a charged ionic layer (charge
stabilization). The present study focusses on the dynamics of
sterically-stabilized suspensions. For such suspensions, the inter-particle
forces are well described by a {\it hard-sphere} interaction potential with no 
detectable long-range interactions and an infinite repulsion when two particle 
centers are separated by one diameter.
\smallbreak
The experimental study of dynamics in dense colloidal suspensions was
pioneered by P. N. Pusey, W. van Megen and collaborators
(see {\em e.g.} \cite{Pusey:Nature1986, VanMegen:JChemPhys85,
VanMegen-Pusey:PRA91, Pusey-chapter-91,Segre_shorttime_PRE95, VanMegen:PRE98}),
using Dynamic Light Scattering (DLS).
The phase behavior of a hard-sphere suspension depends on a single parameter -
the packing fraction, or particle volume fraction, $\Phi$. In the low volume
fraction limit, with $\Phi$ on the order of a few percent or
less, the dynamics of individual particles is essentially Brownian. The
relaxation times measured by DLS yield a $q$-independent diffusion
coefficient $D_0$ equal to that of free particles i.e. the
Stokes-Einstein free diffusion coefficient. However, as soon as the volume
fraction is increased, the dynamics is slowed down by both direct interactions 
between the particles and by indirect hydrodynamic interactions mediated by 
the solvent. These interactions are highly dependent on the structural 
properties of the system and thus on the scattering vector $q$. In order to 
minimize effects introduced by  multiple scattering of light, the DLS studies 
use elaborate refraction index matching procedures between the colloidal 
particles and the solvent or complex scattering techniques such as two-color 
DLS (TCDLS) \cite{Segre:2color:JModOpt95}.
The TCDLS work described in Ref.~\cite{Segre_shorttime_PRE95}
studies the dynamics of hard-sphere colloidal suspensions. The $q$- and
time-dependent short-time diffusion coefficient $D_S(q,t)$ obtained from the
intensity autocorrelation functions could be related, via the static structure
factor $S(q)$, to the {\it hydrodynamic functions} predicted by a many-body
theory derived by Beenakker and Mazur (BM)
\cite{BeenakkerMazur1983,BeenakkerMazur1984}.
The study in Ref.~\cite{Segre_shorttime_PRE95} showed and excellent match 
between TCDLS measurements and the BM predictions in fluid suspensions of
relatively low ($\Phi \leq$ 0.35) volume fractions. On the contrary, at higher
concentrations ($\Phi > 0.4$) the TCDLS data deviated significantly from the 
BM theory.
\smallbreak
In the present study we further investigate these phenomena using a
complementary experimental technique, that is X-ray Photon Correlation
Spectroscopy (XPCS) (for some recent reviews, see
\cite{Grubel-chapter-2008,Sutton:CRPhysique08,Livet:ActaCrystA07} and 
references therein).
XPCS is the equivalent of DLS in the X-ray domain. It is
not affected by problems related to multiple scattering and access to larger
momentum transfers $q$ is possible thanks to the shorter wavelength. An
important consequence of the access to higher $q$ values is that the static
structure factor can be experimentally determined and modeled, {\em e.g.} by 
the Percus--Yevick formalism. XPCS can only be performed at the latest 
generation synchrotron radiation sources where the coherent flux is large 
enough to perform scattering experiments and obtain a good signal-to-noise 
ratio of the correlation functions. Damage to the sample induced by the X-ray 
beam can be a nuisance, especially when studying soft-matter or biological 
systems, and during an experiment one must carefully monitor the state of the 
sample.
\smallbreak
Here we report the results of XPCS experiments on fluid suspensions of 
sterically-stabilized spherical particles with volume fractions up to 
$\Phi\approx$~0.49. Previous XPCS experiments on hydrodynamic effects in 
colloidal suspensions have focused on charge-stabilized particles 
\cite{Lurio_PRL00,Lumma:latex:PRE00,Aymeric:EPJE2008,Riese_screening_PRL00,
banchio_PRL06}.

\section{Experimental details}

\begin{figure}
		\resizebox{0.9\columnwidth}{!}{%
		\includegraphics{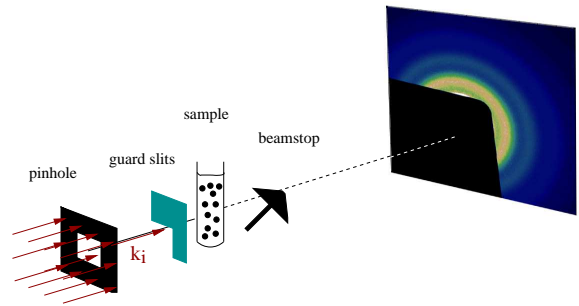}}
        \caption{(Color online) Sketch of the experimental setup for XPCS.}
        \label{fig:XPCS-setup}
\end{figure}

The XPCS experiments were performed using partially coherent X-rays at the
ID10A beamline (Tro\"ika) of the European Synchrotron Radiation Facility (ESRF)
in Grenoble, France. A single bounce Si(111) crystal monochromator was used to
select 8 keV X-rays, having a relative bandwidth of
$\Delta \lambda /\lambda \approx  10^{-4}$.
Higher order light was suppressed by a Si mirror placed in the monochromatic
beam. A transversely partially coherent beam was defined by using a set of
high heat-load secondary slits placed at 33 m from the undulator source,
a beryllium compound refractive lens (CRL) unit placed at 34~m from the
source thereby focusing the beam near the sample location, at 46~m, and by a
set of high precision pinhole slits with highly polished cylindrical edges, 
placed just upstream of the sample, at ~45.5~m 
(see figure~\ref{fig:XPCS-setup}).
The final beam size selected by the beam-defining pinhole slits was
10~x~10~$\mu$m$^2$. The parasitic scattering from the slits was suppressed by a
guard slit placed a few cm upstream of the sample.  Under these conditions,
the partial coherent flux on the sample was $\sim10^{10}$~ph/s.
\smallbreak
The static scattering from the colloidal suspensions was recorded by a
charge-coupled device (CCD) with 22~$\mu$m pixels size located 2.2~m
downstream of the sample. The dynamic information was obtained with a
scintillation detector (Cyberstar) connected to a multiple-tau FLEX01-08D
hardware correlator from \emph{Correlator.com}. The detection area was
limited to a size corresponding to a few speckles by precision slits placed
in front of the point detector. Typical detector slit settings during these
experiments were between 50~x~50~$\mu$m$^2$ and 100~x~100~$\mu$m$^2$.

\begin{figure}
\resizebox{0.8\columnwidth}{!}{%
\includegraphics{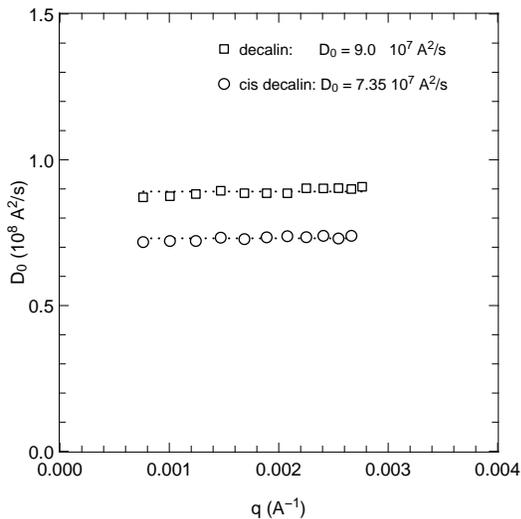}}
\caption{Measurement of D$_0=1/\tau q^2$
        by Dynamic Light Scattering. The squares
	show the ({\it q}-independent) data points obtained with
	the sample (dispersed in decalin) which was used in
	the XPCS experiments. For comparison, data taken from an
	identical sample suspended in cis-decalin (see text) is
	also shown. }
\label{fig:D0}
\end{figure}

The colloidal suspension was prepared by A. Schofield of the University of
Edinburgh and consisted in poly(methyl methacrylate) (PMMA) spherical particles
coated with a thin layer of poly-12-hydroxy steric acid, suspended in decalin.
A net electric charge on the colloidal particles can typically be produced in 
low polarity solvents by the addition of surfactants or charge-control 
additives \cite{Bartlett:La2010}. However, as prepared, the particles suspended 
in decalin are expected to show no detectable traces of residual charges 
(see e.g. \cite{Besseling2009}) and interact as almost perfect hard spheres. 
This assumption is supported by the XPCS measurements presented below which 
are in good agreement with theories assuming a hard sphere interaction 
potential. 
The batch solution had a measured volume fraction of $\Phi$ = 0.327. Higher 
concentration suspensions were obtained using a centrifugation process, while 
lower concentration were prepared by adding more decalin solvent.
\smallbreak
The sample cell consisted in a 1.5~mm diameter Kapton tube with wall thickness
of $\approx$ 80~$\mu$m. A syringe pump purchased from Harvard Apparatus
connected through Teflon tubing and leak tight fittings purchased from
Upchurch Scientific was used to fill the Kapton tube. This experimental setup
has the advantage of allowing measurements under continuous
flow as explored previously \cite{fluerasu:NJP10}. However, all the 
measurements reported here were performed on stationary, non flowing samples.
Our results on the dynamics under continuous flow will be described in a
subsequent publication. The flow option was only used to periodically renew 
the sample by bringing fresh particles into the beam. Through
many repeated measurements it was found that flowing the samples introduces
new time scales in the measured correlation functions
\cite{XPCSflow:JSynch08,Busch:EPJE2008} that, in particular, can affect the 
long-time decays, even for a long time after the flow was stopped. As a
consequence, a stabilization time was always allowed after renewing the
samples or filling the flow cell.
\smallbreak
In the high-dilution limit, hydrodynamic or direct interactions between the 
particles are negligible and the colloids undergo Brownian motion with a 
diffusion coefficient $D_0$ described by the Stokes-Einstein diffusion 
relationship
\begin{equation}
D_0 = \frac{k_B T}{6 \pi \eta a_H},
\label{eq:ES}
\end{equation}
where $\eta$ is the viscosity and $a_H$ the hydrodynamic radius.
$D_0$ was measured by DLS on samples with $\Phi < 1 \%$. It should be mentioned
that with this sample concentration, it is impossible to perform XPCS
measurements due to the weak scattering of X-rays. The DLS measurements were 
performed at several scattering angles and wavelengths of $\lambda$=532~nm and 
633~nm. The time constants $\tau$ were obtained by fitting the correlation 
functions with simple exponentials, and the momentum transfer $q$ was 
calculated from the scattering angle $2\theta$ using
\begin{equation}
q=\frac{4 \pi n}{\lambda}\sin \frac{2\theta}{2}.
\end{equation}
Here $n$=1.48 is the index of refraction of the solvent - decalin, a mixture of
50/50 cis- and trans-decalin as determined using the viscosity
measurements described below \cite{cis_visco,cis_n}.
According to Fick's law, the mean square displacement of Brownian particles
from their position at $t$=0 is $<\Delta x^2>=6D_0t$ with $D_0$ being the 
diffusion coefficient. In reciprocal space, this corresponds to a 
$q^{-2}$ dependence of the correlation time $\tau$,
and a diffusion coefficient $D_0=1/(\tau q^2)$. 
As seen if figure~\ref{fig:D0} this quantity is 
independent of the scattering angle, leading to a diffusion 
coefficient for the PMMA particles in decalin of 
$D_0$=9$\times 10^7$~\AA $^2$/s.

As a cross-check, $D_0$ was measured also for a second sample, consisting of
the same PMMA particles suspended in pure cis-decalin, which has a nominal
viscosity of 3.06~cP (at 24 $^\circ$C, \cite{cis_visco}). The results of the 
DLS measurements, are summarized in Table~\ref{table:D0}. The viscosity
of the dilute suspensions was directly measured for both samples using a 
U-tube viscometer, hence allowing to determine $a_H$ from equation~\ref{eq:ES}.
The hydrodynamic radii $a_H$ measured in the two different solvents are
equal within the error bars.

\begin{table}

\caption{\label{tab:D0}Measured $D_0$ (DLS) and viscosity $\eta$
(U tube viscometer) for low concentration suspensions of PMMA
particles in decalin and cis-decalin (at $24^\circ$C), and the
hydrodynamic radius $a_H$ calculated from the Stokes-Einstein
relationship}
   \begin{ruledtabular}
      \begin{tabular}{|c|c|c|c|}
      \textbf{solvent}& $\eta$ (cP) & $D_0 (\text{\AA}^2 /s)$ & $a_H (\text{\AA})$ \\
      decalin 	& $2.6 \pm 0.1$	& $(9.0 \pm 0.2) \times 10^7$  & $931\pm59$\\
      cis-decalin & $3.2\pm0.1$ & $(7.35\pm0.07)\times 10^7$  & $925 \pm 38$\\
   \end{tabular}
   \end{ruledtabular}
   \label{table:D0}
\end{table}

\section{Results}

\subsection{Static properties}
\begin{figure}
  \resizebox{0.9\columnwidth}{!}{%
  \includegraphics{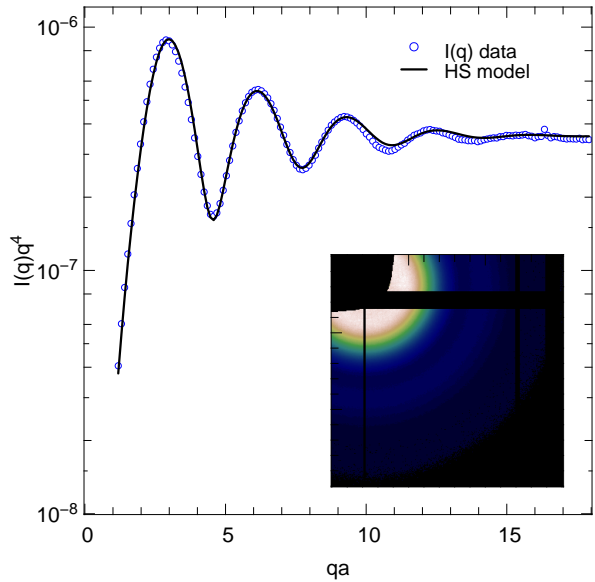}}
  \caption{(Color online) Example fit for the static scattering from the
	   $\Phi=$18.5~\% sample. The SAXS signal (plotted here as
	   $I(q) q^4$ to emphasize the agreement with Porod's
	   law) was obtained by circular averaging the 2D CCD images
	   (inset). The continuous line shows a fit with a model assuming
	   a polydisperse suspension of uniform spheres interacting via a hard
           sphere repulsive potential, as described in the text.}
\label{fig:Iq}
\end{figure}

SAXS measurements, performed at the ID10A beamlines (ESRF), were
corrected for background scattering contributions from the solvent, the Kapton 
tubes of the sample environment, etc. The static data was fitted using the
Percus--Yevick (PY) closure \cite{Griffith:Sq:PRA86} for the structure factor
$S(q)$, and a form factor $P(q)$ for spherical particles obtained from the
fits on the lower concentration suspensions. The expressions for $S(q)$ and 
$P(q)$ are calculated assuming a particle size polydispersity described by the 
Schultz distribution function \cite{KotlarchykChen:JChemPhys83}. In addition, 
the resulting form for the scattered intensity $I(q) \propto P(q) S(q)$ 
was convoluted with a Gaussian function describing the instrumental resolution.
An example of the resulting fits for $I(q)$ is shown in figure~\ref{fig:Iq}
for the $\Phi = 18.5\%$ sample. The same procedure was applied for all samples. 
The fitting parameters are the particle radius $a$=890$\pm$12~\text{\AA}~with 
a size distribution standard deviation $\sigma _a$=89$\pm$11~\AA, the volume 
fraction $\Phi$ of each individual suspension, and an overall multiplicative 
factor measuring the scattering cross section of each sample which is not 
discussed here. On a subset of the samples, additional SAXS 
measurements were performed on the SAXS beamline ID02 at ESRF. The data 
measured on the two different instruments (ID10 and ID02) are in excellent 
agreement.

\begin{figure}
  \resizebox{\columnwidth}{!}{%
  \includegraphics{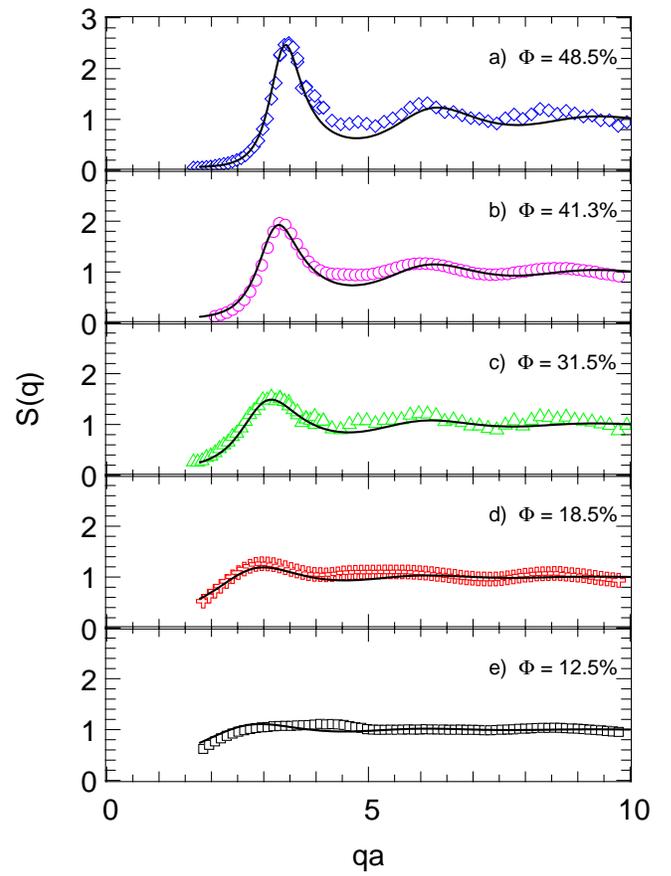}}
  \caption{(Color online) Structure factor $S(q)$ for various samples. Data points are
  calculated, as described in the text, from the measured scattered intensity
  and the polydisperse particle form factor resulted from the SAXS fits. Error
  bars are estimated to be smaller than the symbols. The solid lines show the 
  fitted structure factor with the PY model for a polydisperse suspension.}
\label{fig:Sq}
\end{figure}

The PY SAXS analysis procedure described above provides a fitted form factor 
for the PMMA particles and a fitted structure factor for each of the 
suspensions. An ``experimental structure'' factor is not directly accessible 
from the data but, assuming that the decoupling approximation works well
for the relatively monodiperse suspension of spherical particles studied here 
\cite{SkovPedersen97}, a good estimate for it can be obtained by dividing the 
experimental scattered intensity $I(q)$ with the fitted form factor,
\begin{equation}
S(q) \propto I(q)/P(q).
\label{eq:Sqexp}
\end{equation}
The structure factors $S(q)$ for several samples with different concentrations
can be seen in figure~\ref{fig:Sq}, where the fitted $S(q)$ (continuous lines)
are shown together with the ones calculated form the experimental data points
using equation \ref{eq:Sqexp}. The fitted volume fraction is indicated on the 
graph for each of the individual samples. The agreement is very good for all
concentrations and over the whole $q$-range. We attribute the small 
discrepancies  that appear in some of the fits to experimental artifacts
such as parasitic scattering. 

\subsection{Dynamic behavior}

The intensity fluctuation autocorrelation functions
\begin{equation}
  g^{(2)}(q,t) = \frac{\left< I(q,t_0) I(q,t_0+t) \right>_{t_0}}
                      {\left< I(q,t_0)\right>_{t_0}^2},
\label{eq:g2}
\end{equation}
where measured for wave vectors $q$ around the main peak in the structure factor
$S(q)$ in a range of 1.5$\leq qa \leq$6.

\begin{figure}
  \resizebox{\columnwidth}{!}{%
  \includegraphics{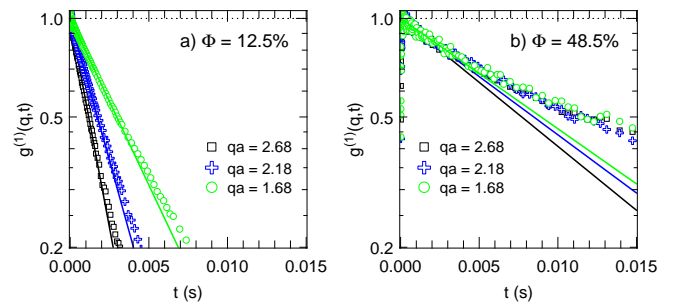}}
  \caption{(Color online) Intermediate scattering function $g^{(1)}(q,t)$, 
            measured at
           $\Phi \simeq 12.5~\%$ and at $\Phi \simeq 48.5~\%$ for several q 
           values around the structure peak. Solid lines show fits with 
           equation \ref{eq:g1-D0} for the low concentration sample and
           fits for the initial decay of the correlation functions measured
           from the high concentration suspension using equation \ref{eq:g1-Ds}
	   as described in the text.}
\label{fig:g1}
\end{figure}

Assuming a Gaussian distribution of the temporal fluctuations at a fixed $q$,
the normalized dynamic structure factor, or intermediate scattering function
(ISF),
\begin{equation}
g^{(1)}(q,t)=\left| \frac{S(q,t)}{S(q,0)} \right|,
\label{eq:g1}
\end{equation}
is related to the intensity autocorrelation functions via the Siegert
relationship,
\begin{equation}
g^{(2)}(q,t)=1+\beta \left[ g^{(1)}(q,t) \right]^2.
\label{eq:siegert}
\end{equation}
Here, $\beta$ is the optical contrast, which depends on the transverse
coherence lengths of the X-ray source and the sample, and on geometrical
parameters such as the pinhole size and the detector slit opening. In the 
experiments described here, the contrast $\beta$ was around 2-5\%.

In low concentration samples, the ISF measured at a wave vector $q$ is a
simple exponential decay, with the relaxation rate depending on the single
particle diffusion coefficient $D_0$ and $q$
\begin{equation}
g^{(1)}(q,t)=\exp \left[-D_0 q^2 t \right].
\label{eq:g1-D0}
\end{equation}

For higher concentration suspensions, both direct interactions (DI) acting 
via the hard-sphere interparticle potential and 
hydrodynamic interactions (HI) between colloids mediated by the solvent 
start playing an increasingly important role, slowing down the diffusive 
dynamics of the particles. While the quasi-instantaneous HI between
particles are related to the structure factor $S(q)$ they do not,
themselves, determine the equilibrium static structure, nor do they shift the
glass transition concentration (see {\em e.g.} \cite{FuchsMayr:PRE99}).
However, the effects of the HI are very important, even at relatively low
volume fraction as they slow down considerably the short-time relaxations.
Our XPCS results on this complex many-body process are shown in the 
following sections.

\subsubsection{Short-time dynamics and Hydrodynamic Interactions}

An important time scale arising in dense colloidal suspensions is the
{\it short time} limit $\tau _S$. This is usually associated with the random
motion of individual particles in cages formed by neighboring particles.
For $t>\tau _S$ the diffusive motion is slowed down by both
HI and DI between the particles.  At $t<\tau _S$, the motion is still 
diffusive but faster, as it is slowed down only by HI and not yet by DI. This 
effect can be clearly seen in figure~\ref{fig:g1}. The ISFs are plotted here 
for three different values of $q$ and two different samples - a 
low-concentration one, $\Phi$=12.5~\%, and a high concentration one - 
$\Phi$=48.5~\%. 

While the low concentration sample shows single exponential decays, in the 
high concentration samples two distinct relaxation rates are observed. On the 
semi-logarithmic scale used in figure~\ref{fig:g1}, the exponential ISFs 
$g^{(1)}(q,t)$ appear as straight lines, with the relaxation rates measured 
by their slopes.
A short-time of $\tau _S~\approx$~2~ms can be estimated from the change in
slope obvious with the high concentration sample (but absent in the low
concentration one). A physical interpretation of this time scale emerging
in high density suspensions, was given by Segr\'e, Behrend, and
Pusey~\cite{Segre_shorttime_PRE95}.
There, $\tau_S$ is the time required for a particle to diffuse away from the 
position at $t$=0 to a distance equal to its radius.
With our experimental parameters, this leads to $\tau _S=a^2/D_0\approx 9~$ms. 
While this is within the same order of magnitude with the value observed 
experimentally, the agreement is clearly not very good. An obvious problem
associated with this estimate for $\tau _S$ is associated with the fact that
the free diffusion coefficient $D_0$ was used while the diffusion is known to be
slowed down by HI. Replacing $D_0$ with a slower diffusion coefficient
(e.g. $D_S(q)$ - see discussion below) makes the disagreement between the 
experimentally observed $\tau _S$ and the calculated one even stronger.
This is, however, not surprising, because the length scale over which the 
particles can move before being affected by DI should be dependent on the 
volume fraction, and is not necessarily equal to the particle radius.  
A different estimate of the short time $\tau _S$ can be achieved using an 
expression for the frequency of collisions derived from a theory by 
M.~Smoluchowsky which is usually used to describe coagulation kinetics 
(although here we assume that coagulation is prevented by the steric 
stabilization of the particles).
For a suspension of $N$ colloidal particles in a solvent of dynamics viscosity 
$\eta$ and a total volume $V$, the rate of collisions is given by (see for e.g. 
\cite{Holtfoff:Langmuir96,Lin:JPCM90} and references therein),
\begin{equation}
k=\frac{8k_B T}{\eta}\frac{N}{V}.
\label{eq:l}
\end{equation}
With a particle radius $a$, and volume fraction $\Phi$, this results in a time 
between collisions, identified here with $\tau _S$ of
\begin{equation}
\tau _S=\frac{1}{k}=\frac{\pi \eta a_H^3}{6k_B T \Phi}.
\label{eq:taus}
\end{equation}
For $\Phi$=48.5~\%, and using the values of $a_H$ and $\eta$ from 
Table~\ref{table:D0}, equation~\ref{eq:taus} leads to a short time of
$\tau_s \approx 0.5$~ms. Since even on these short time scales the dynamics
is considerably slowed down by HI the actual time is expected to be slower.
Considering a factor of $\approx$10 for this slowing down, in agreement
with data shown further below for the 48.5~\% sample, a time scale of 
$\tau_S \approx$~5~ms can be estimated, which is in better agreement with the 
data in Figure~\ref{fig:g1}.

\begin{figure}
  \resizebox{\columnwidth}{!}{%
  \includegraphics{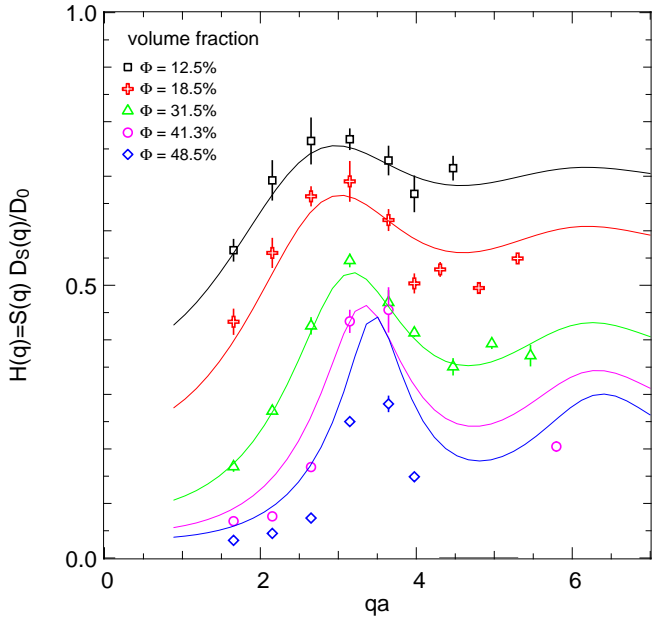}}
  \caption{(Color online) Hydrodynamic functions $H(q)$ vs $qa$. The data 
points are extracted from the fitted short-time diffusion coefficient $D_s(q)$ 
and the static
structure factors $S(q)$, using equation~\ref{eq:Ds}. The volume fractions for 
the different suspensions are indicated on the graph. The solid lines are 
theoretical predictions of the $\delta$-$\gamma$ expansion
\cite{BeenakkerMazur1983,BeenakkerMazur1984}.}
\label{fig:Hq}
\end{figure}

In the short-time limit, the ISFs can be described in terms of a $q$-dependent
diffusion coefficient \cite{Pusey-chapter-91},
\begin{equation}
g^{(1)}(q,t)=\exp\left[-D_S(q) q^2 t \right],
\label{eq:g1-Ds}
\end{equation}
with the (short-time) diffusion coefficient given by,
\begin{equation}
D_S(q)=D_0 \frac{H(q)}{S(q)}.
\label{eq:Ds}
\end{equation}
Here, the hydrodynamic function $H(q)$ describes the HI. In the high dilution
limit, $S(q)=1$ and $H(q)=1$, leading to $D_S(q)=D_0$. A non-unitary
hydrodynamic function, $H(q) \neq 1$ is a hallmark of HI. The short--time
diffusion coefficients were determined by a first cumulant analysis
\cite{Segre_physicaA97} in which the initial decay of the correlation function
was fitted to an exponential form. The fits were performed in two stages.
First, the correlation functions are fitted over the entire time range with a
stretched exponential form,
\begin{equation}
g^{(2)}(q,t)=\beta \exp\left[ -2\left( \Gamma t\right) ^\gamma \right] +g_\infty,
\label{eq:stretchedexp}
\end{equation}
to obtain accurate values for the experimental contrast $\beta$
and baseline $g_ \infty$ (with $g_\infty\approx1$ for all the correlation
functions). Subsequently, these parameters are being fixed in a second fit, 
performed only for the initial decays $t<t_s$, with a simple exponential form,
\begin{equation}
g^{(2)}(q,t)=\beta \exp\left[-2D_s(q)q^2t\right] +g_\infty,
\label{eq:g2-Ds}
\end{equation}
where the $q$-dependent short-time diffusion coefficient $D_s(q)$ is the only
free parameter. In these ergodic systems where the intensity
fluctuations are well described by Gaussian statistics, equations 
\ref{eq:g2-Ds} and \ref{eq:g1-Ds} are equivalent.

The $\delta$-$\gamma$ expansion proposed by the BM theory
\cite{BeenakkerMazur1984} provides one of the most successful tools to date
describing HI in dense but fluid colloidal suspensions. The only input 
parameter required is the static structure factor $S(q)$. 
The BM theory predictions were verified by XPCS in several different 
charge-stabilized suspensions with 
screened electrostatic interactions \cite{banchio_PRL06, Aymeric:EPJE2008}.
In the case of sterically-stabilized suspensions, interacting via a
hard-sphere potential, the predictions of BM theory where verified by the
two-color DLS experiments described in \cite{Segre_shorttime_PRE95}.
One problem associated with using light scattering is that in general it is
impossible to reach high enough values of the scattering vector $q$ to obtain
accurate structural information, hence calculated values for $S(q)$
(using the PY closure) are usually taken as input.
In X-ray scattering experiments, access to $S(q)$ down to a fraction
of the colloidal length scale is straightforward so here the hydrodynamic 
functions $H(q)$ were calculated
from equation~\ref{eq:Ds} using measured static and dynamic data.
One of the problems encountered in XPCS experiments is that accurate 
dynamic data can only be obtained in relatively small $q$-range ({\em e.g.} 
compared to DLS) because the signal-to-noise is often limited by the 
decreasing scattering cross section at high $q$ (or, equivalently, by a 
limited intensity of the coherent X-ray beam). The results for $H(q)$ can be 
seen in figure~\ref{fig:Hq} for a wide range of volume concentrations and a 
range of $q$ covering
2.14$\times 10^{-3}$\AA$^{-1}\leq$~$q$~$\leq$6.44$\times 10^{-3}$\AA$^{-1}$
(or 2$\leq qa \leq$6). Experimental data measured at different concentrations
are shown by the different symbols specified in the legend. Continuous lines
show predictions of the $\delta$-$\gamma$ expansion with no adjustable 
parameters other than the corresponding volume fractions resulted from the
$S(q)$ fits. The agreement between theory and experiment is
excellent for solutions with $\Phi \le$~0.4 which is well in line with the
earlier observations in Ref.~\cite{Segre_shorttime_PRE95}. The
$\delta$-$\gamma$ expansion employed in the calculation is expected to break
down for suspensions at high volume fractions
\cite{BeenakkerMazur1984}. This is indeed confirmed by our data, which
shows a clear overestimate of $H(q)$ at $\Phi$=48.5~\%.

\begin{figure}
  \resizebox{0.8\columnwidth}{!}{%
  \centering \includegraphics{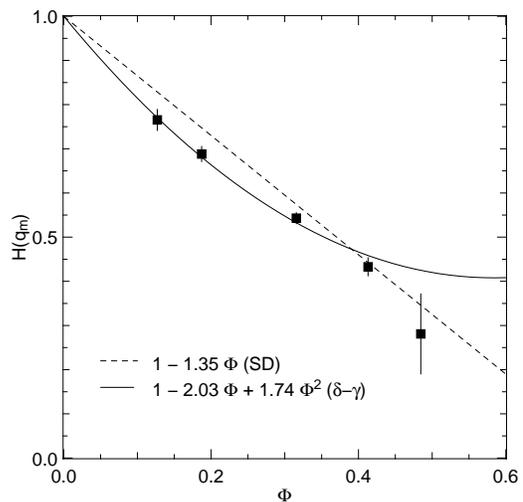} }
  \caption{
Comparison between $H(q_m)$ and polynomial analytic forms that fit the
theoretical predictions by the $\delta$-$\gamma$ theory (continuous solid line)
and the Stokesian dynamics numerical algorithm by Banchio {\em et al.} (dashed
straight line).}
\label{fig:Hqm-dg-SD}
\end{figure}

\begin{figure}
  \resizebox{0.85\columnwidth}{!}{%
  \centering \includegraphics{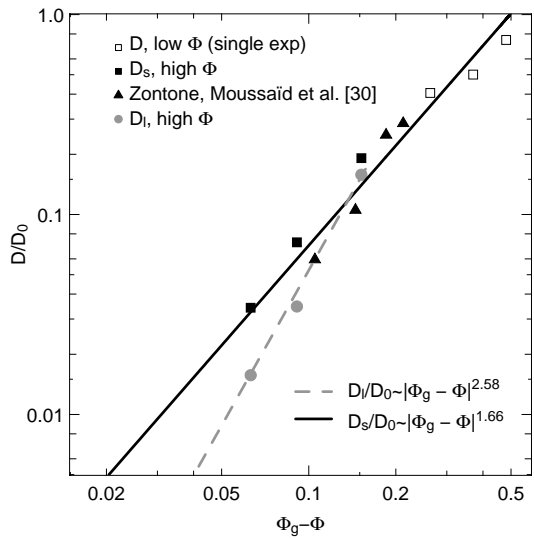} }
  \caption{ Normalized short- and long--time diffusion coefficients 
  $D_s(q_m)/D_0$, $D_l(q_m)/D_0$ measured at the peak of $S(q)$ ($q=q_m$) versus 
  the separation parameter $\left|\Phi-\Phi_g\right|$. The continuous and 
  dashed straight lines are mode-coupling theory predictions as described in 
  the text.}
\label{fig:D-Phi-scaling}
\end{figure}

Figure~\ref{fig:Hqm-dg-SD} shows the value of the Hydrodynamic function near 
the structure factor maximum $H(q_m)$ as a function of the volume fraction. 
Within the $\delta$-$\gamma$ expansion, the $\Phi$-dependence of the peak 
value of $H(q_m)$ for hard spheres with PY input for $S(q)$ and $\Phi\leq0.45$
is well parametrized by a quadratic form 
\begin{equation}
H(q_m)=1-2.03\Phi+1.74\Phi^2,
\label{eq:Hqm-dg}
\end{equation}
represented by the continuous solid line in figure~\ref{fig:Hqm-dg-SD}.
The results are also compared with predictions of numerical calculations
using the Accelerated Stokesian Dynamics (ASD) algorithm by Banchio {\em et al.}
\cite{Banchio-Nagele:JChemPhys08,banchio:JCP99} for the HI of hard spheres.
Their numerical results predict a linear dependence of $H(q_m)$ as a function
of $\Phi$ well approximated by
\begin{equation}
H(q_m)=1-1.35\Phi,
\label{eq:Hqm-SD}
\end{equation}
which is represented by the dashed straight line in figure~\ref{fig:Hqm-dg-SD}. 
As it can be seen, the ASD numerical simulations provide good predictions 
over the entire range of concentrations and really excellent predictions 
above $\Phi\ge$0.4 where the $\delta$-$\gamma$ theory fails. This is in 
agreement with the results reported in \cite{banchio_PRL06} for 
charged-stabilized particles with screened electrostatic interactions.

The dynamic of fluctuations in suspensions with increasing volume fraction,
is also expected the be well described by the Mode Coupling Theory (MCT) of the
glass transition in hard-sphere colloids
\cite{GotzeSjogren:PRA91, VanMegen:PRE07}. In its lowest order in the
separation parameter $\left| \Phi _g-\Phi \right|$, the MCT
predicts the existence of two divergent time scales describing the dynamics in
different windows of time - the $\beta$-relaxation at fast time scales and
the $\alpha$-relaxation at slower time scales. The short-time diffusion 
coefficient $D_s(q_m)$ corresponding to the fast $\beta$-relaxation 
measured near $q=q_m$ follows a power scaling law,
\begin{equation}
D_s(q_m) \propto \left|\Phi _g - \Phi \right| ^{1.66}.
\label{eq:DsPhi}
\end{equation}
This form is valid both in the liquid, $\Phi < \Phi _g$ (the systems
probed here) or glass, $\Phi > \Phi _g$ states \cite{GotzeSjogren:PRA91}. The
fundamentally different $\alpha$-process, which is completely frozen in the 
glassy state, restores ergodicity in the high concentration liquid phase 
with a characteristic long-time diffusion coefficient $D_l(q)$ following
\begin{equation}
D_l(q_m) \propto \left|\Phi _g - \Phi \right| ^{2.58}.
\label{eq:DlPhi}
\end{equation}
The slow relaxation process in the high density fluids described here is 
analyzed in more detail in the following section. Here we 
investigate the scaling forms proposed by the MCT - 
equations~\ref{eq:DsPhi}, \ref{eq:DlPhi} - for the diverging short-time and 
long-time relaxations. The results can be seen in 
figure~\ref{fig:D-Phi-scaling}. 
For the low density suspensions, the diffusion coefficients obtained from 
single exponential fits are shown by the open black squares. The short-time 
diffusion coefficients obtained from the high density suspensions using the 
first-cumulant analysis described above are shown by the solid black 
squares. We also show additional data (solid black triangles) for the 
short time diffusion coefficient previously obtained by 
Zontone, Moussa\"id {\em et al.} \cite{Abdellatif_colloid} in a very similar 
system - PMMA hard sphere particles suspended in {\it cis-}decalin. 
All the data follows well the scaling law predicted by the MCT and 
represented by the black continuous line in figure~\ref{fig:D-Phi-scaling} 
with $\Phi_g=0.585$.

To date, there are fewer points available for the long-time diffusion 
coefficients. These are shown in figure~\ref{fig:D-Phi-scaling} by the grey 
solid circles, which are also in good agreement with the MCT
scaling form (grey dashed power law). The slow diffusion 
coefficients were obtained from the fits
shown in figure~\ref{fig:g1-highPhi}, and described in the following section.

The experiments presented in this section provide further evidence for two
important results on the dynamics in high density colloidal suspensions
interacting with a hard-sphere potential:

{\it i)} At moderate concentrations $\Phi\ll\Phi _g$ (i.e. $\Phi\leq$~40~\%) 
the short time diffusion coefficients and HI are well described quantitatively 
by the BM theory. At higher concentrations, only the ASD numerical results by
Banchio {\em et al.} provide accurate results.

{\it ii)} The MCT of the colloidal glass transition provides correct accurate
quantitative predictions for both the short-time and the long-time diffusion
coefficients in colloidal suspensions of higher concentrations
($\Phi \leq \Phi _g$).

\subsubsection{Long--time and short--time behavior}

\begin{figure}
  \resizebox{0.9\columnwidth}{!}{%
  \includegraphics{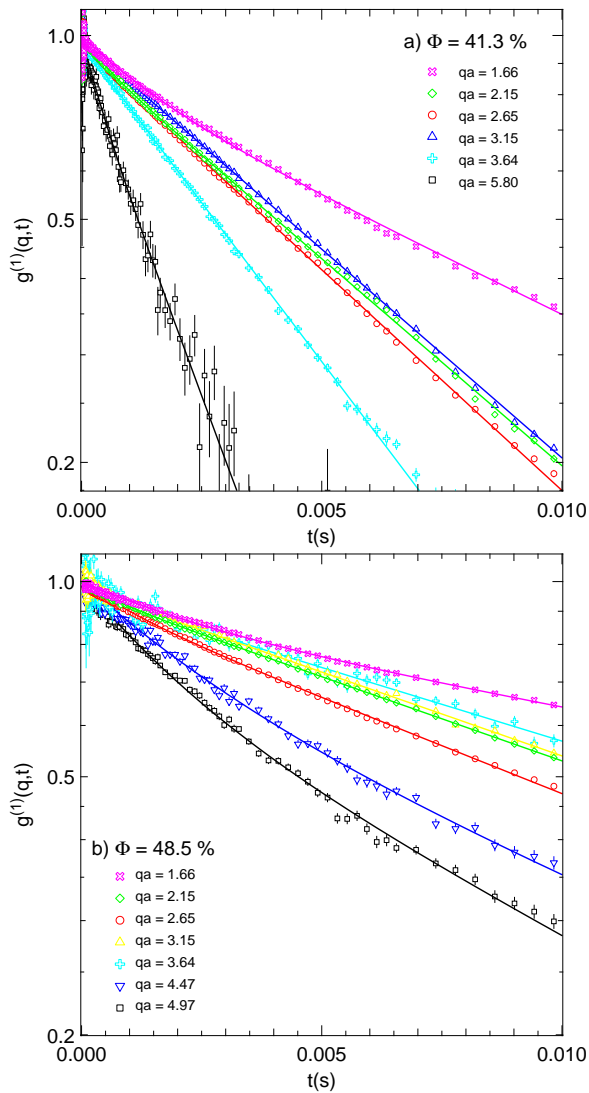}}
  \caption{(Color online) Intermediate scattering function $g^{(1)}(q,t)$ 
           measured at $\Phi \simeq$ 41.3~\% (a) and 48.5~\% (b) for several 
           $q$ values around the structure peak and fits with double exponential
           decays (continuous lines).}
\label{fig:g1-highPhi}
\end{figure}

\begin{figure}
  \resizebox{0.9\columnwidth}{!}{%
      \includegraphics{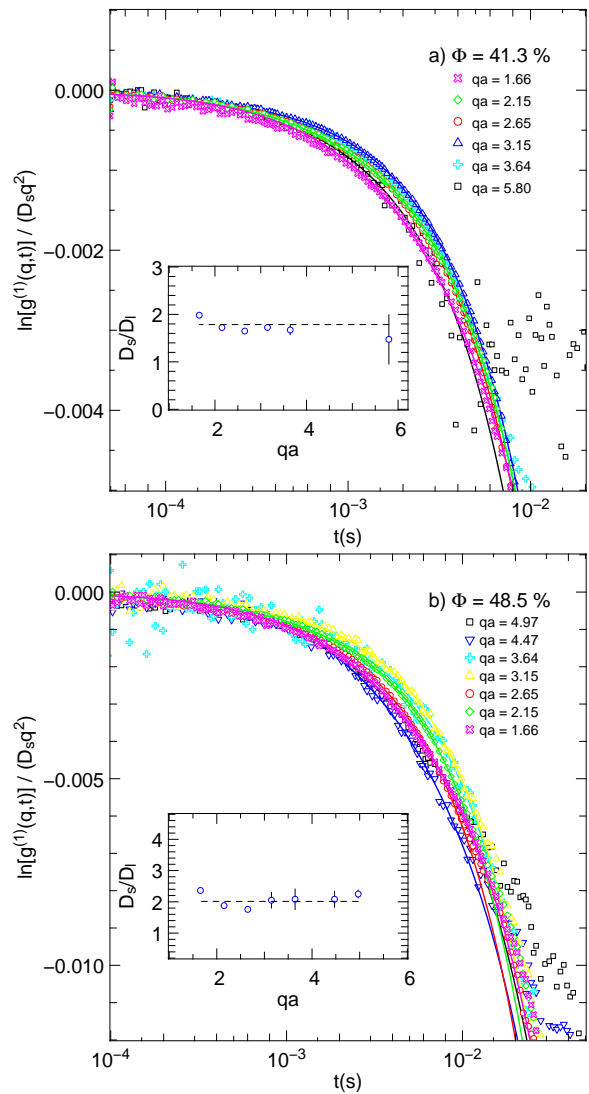}}
  \caption{(Color online) Intermediate scattering functions for all values of 
$q$ measured (same data as in figure~\ref{fig:g1-highPhi})
scaled by the corresponding short-time relaxation rates $\Gamma_s=D_sq^2$ at
$\Phi$= 41.3~\% (a) and 48.5~\% (b).
The insets show the ratio between the fitted short-time and long-time
diffusion constants versus $qa$. For both samples this ratio is, within our
experimental accuracy, a $q$-independent constant.}
\label{fig:g1_scaling}
\end{figure}

In this section we take advantage of the fact that both the short-time and
long-time diffusion coefficients are readily available from the XPCS data,
to test an approximate scaling law, first proposed by
Segr\'e and Pusey~\cite{Pusey_scalingSq_PRL96}. By using the 
aforementioned TCDLS technique, they evidenced a proportionality between the
short- and long-time diffusion coefficients measured in concentrated
suspensions with $\Phi \approx$0.46 and higher, over a broad range of $q$ 
values (excluding the smallest $q$s). 
This proportionality results in a collapse on a single master curve of
the entire intermediate scattering functions measured at different $q$ values 
when scaled by their short-time decays. This finding suggests that the 
structural relaxations of particles or ``cages of particles'' are both related 
to self-diffusion, which contradicts a  MCT picture where the
$\alpha$- and $\beta$-relaxations have different physical origins.
However, subsequent MCT results \cite{FuchsMayr:PRE99} provided a
semi-quantitative argument for the approximate collapse of the ISFs on a single master curve at high-enough values of $q$ and low-enough values of
time, where the non-diffusive character of the $\alpha$-relaxations is not
very pronounced and the two modes hence follow similar decays. The MCT 
results also point to the fact that this scaling form is not expected to be 
valid in charge-stabilized suspensions due to the long-range repulsive nature 
of the potential. This was indeed observed experimentally in previous XPCS 
experiments by Lurio, Lumma, Mochrie {\em et al.}
\cite{Lurio_PRL00, Lumma:latex:PRE00}. Interestingly, a recent study 
by Martinez {\em et al.}~\cite{martinez:JCP11} using XPCS and DLS did not 
manage to unambigously detect the long-time dynamics away from the peak of 
$S(q)$ in high concentration suspensions of PMMA particles, and hence could 
not validate the Segr\'e-Pusey scaling. However, their study, together
with an earlier experiment by Riese {\em et al.} \cite{Riese_screening_PRL00}
provide a clear experimental proof of the equivalence between XPCS and DLS 
measurements. 
\smallbreak
The results presented here are obtained on a system very similar to that
used by Segr\'e and Pusey, or more recently by Martinez {\em et al.}
\cite{martinez:JCP11}, namely sterically stabilized PMMA uniform spheres 
(albeit about 1/2 the size), and by use of XPCS.
As pointed out before, with the smaller particles, the higher concentration 
suspensions show clearly two different relaxation time scales within the time
and length scales accessible in the XPCS experiments. In order to extract them 
from the experimental data, the ISFs are fitted with double exponential 
decays
\begin{equation}
g^{(1)}(q,t)=A \exp (-\Gamma_1 t)+(1-A) \exp (-\Gamma_2 t).
\label{eq:dexp-fits}
\end{equation}

The fits for the two high concentration suspensions studied here with
$\Phi$=41.3 \% and $\Phi$=48.5~\% can be seen in figure~\ref{fig:g1-highPhi}.
From this procedure, both $D_s(q)$ and $D_l(q)$, the short-time and long-time
diffusion coefficients, associated with the MCT $\beta$- and 
$\alpha$-relaxations, respectively, are readily available.

In order to test the Segr\'e-Pusey scaling relationship between $D_s$ and
$D_l$, $\text{ln} \left[ g^{(1)}(q,t)/(D_sq^2) \right]$ are plotted in figure
\ref{fig:g1_scaling} as a function of time like in the original reference
\cite{Pusey_scalingSq_PRL96}. These are the exactly the same correlation 
functions as shown in figure~\ref{fig:g1-highPhi}, except that the above 
scaling has been performed and a log-lin scale is used.
As it can be seen, the correlation functions measured over more than 
two decades in time collapse on a single master curve, in agreement
with the scaling proposed by Segr\'e and Pusey.
The insets in figure~\ref{fig:g1_scaling} show the ratios between the short-
and the long-time diffusion coefficients $(D_s/D_l)$ at several values of 
$q$. As expected according to the putative scaling behavior this ratio is a 
$q$-independent constant within the experimental error bars.

We conclude that the Segr\'e-Pusey scaling behavior is validated within the 
combination of length- and time scales accessed here, for dense colloidal 
suspensions with $\Phi \approx$40-50~\% interacting via a hard-sphere 
potential. 

\section{Conclusions}

In summary, we have used a combination of XPCS, SAXS and DLS data to measure
the $q$-dependent short time diffusion coefficients $D_s(q)$ and
the hydrodynamics interaction functions $H(q)$ in dense colloidal suspensions
with a hard-sphere interaction potential. Our results show good
agreement with the BM analytical theory at moderate volume
fractions ($\Phi <$0.4). The XPCS data shows good agreement with the ASD
numerical scheme proposed by Banchio {\em et al.} over the entire concentration 
range, and in particular above $\Phi \approx$40~\% where the BM theory is less 
accurate.

The short time diffusion coefficients measured at different volume fractions
scale with the separation parameter $\Phi-\Phi _g$ as predicted by the 
Mode Coupling Theory. 

Finally, the XPCS results for the short-time and long-time dynamics in
high density suspensions are in good agreement with the scaling relationship
proposed by Segr\'e and Pusey.

We wish to acknowledge the ESRF ID10 and ID02 beamlines for providing the 
beamtime, T.~Narayanan for help with the ID02 SAXS experiments and 
A.~Schofield for the sample preparation. The work at Brookhaven 
National Laboratory was performed under Contract No. DE-AC02-98CH10886 with 
the U.S. Department of Energy.

%-----------------------------------------------
% REFERENCES
%-----------------------------------------------

%\bibliographystyle{apsrev}
\bibliography{references}
\end{document}